\documentclass[12pt,a4paper]{article}
\usepackage{amsmath}    %
\usepackage{amsfonts}   %
\usepackage{amssymb}    %
\usepackage{natbib}
\usepackage{rotate}
\usepackage{lscape}
\usepackage[flushleft]{threeparttable}
\usepackage{longtable}
\usepackage{colortbl}
\usepackage{supertabular}
\usepackage{mathrsfs}
\usepackage[brazil,english]{babel} 
\usepackage[latin1]{inputenc}   
\usepackage[T1]{fontenc}        
\usepackage{graphicx}
\DeclareGraphicsExtensions{.pdf,.gif,.jpg, ps, eps, png}

\usepackage{flafter}    
\usepackage{placeins}   

\usepackage[onehalfspacing]{setspace}   
\usepackage{geometry}
\newcommand{\bz}{\mathbf{z}}

\newcommand{\bTheta}{\mbox{\boldmath $\Theta$}}

\newcommand{\bxi}{\mbox{\boldmath $\xi$}}
\newcommand{\bphi}{\mbox{\boldmath $\phi$}} 
\newcommand{\bpsi}{\mbox{\boldmath $\psi$}}

\RequirePackage{hyperref}

\usepackage{tablefootnote}

\begin{document}



\title{A fully likelihood-based approach to model survival data with crossing survival curves}
\author{Fábio Nogueira Demarqui\thanks{Departamento de Estatística, Universidade Federal de Minas Gerais, email: fndemarqui@est.ufmg.br}  \\ Vinicius Diniz Mayrink\thanks{Departamento de Estatística, Universidade Federal de Minas Gerais, email: vdm@est.ufmg.br} }

\date{}

\maketitle

\begin{abstract}
Proportional hazards (PH), proportional odds (PO) and accelerated failure time (AFT) models have been widely used to deal with survival data in different fields of knowledge. Despite their popularity, such models are not suitable to handle survival data with crossing survival curves. Yang and Prentice (2005) proposed a semiparametric two-sample approach, denoted here as the YP model, allowing the analysis of crossing survival curves and including the PH and PO configurations as particular cases. In a general regression setting, the present work proposes a fully likelihood-based approach to fit the YP model. The main idea is to model the baseline hazard via the piecewise exponential (PE) distribution. The approach shares the flexibility of the semiparametric models and the tractability of the parametric representations. An extensive simulation study is developed to evaluate the performance of the proposed model. In addition, we demonstrate how useful is the new method through the analysis of survival times related to patients enrolled in a cancer clinical trial. The simulation results indicate that our model performs well for moderate sample sizes in the general regression setting. A superior performance is also observed with respect to the original YP model designed for the two-sample scenario.
\end{abstract}


\section{Introduction} \label{sec_intro}

Proportional hazards (PH) models have played a central role in the analysis of survival data. Such class of models provides a very flexible framework to model survival data. They further allow an easy interpretation of the parameters from the practical point of view. The main assumption of the PH models is the proportionality of the hazard ratios over time. When such assumption is not verified by the data, some alternatives such as the proportional odds (PO) and the accelerated failure time (AFT) models can be used in the analysis. However, none of them is suitable to accommodate survival data with crossing survival curves. This type of problem is often related to studies involving treatment and control groups. The survival function for one group may have a fast decay in contrast with a slow decay for the other. The curves tend to intersect at some time point configuring an inversion in terms of who is on the top/bottom position. Studying this alteration is relevant in many clinical trials, where the identification of the crossing time indicates when the target treatment for a disease can be considered effective.    

Survival data with crossing survival curves may arise due to several reasons in practice. For instance, \cite{2013_Diao} indicates that this may occur in certain clinical trials related to aggressive treatments such as surgery. Some adverse effects can be observed in an initial stage, but beneficial results may appear in the long run. According to \cite{Bres74}, another situation connected with crossing survival functions is when a treatment has an early and quick effect and it becomes similar to or worse than the placebo treatment after certain period.

Several approaches have been proposed in the literature to accommodate this crossing feature in survival data. The most popular ones are based on time-varying regression coefficients; see, for example, the references \cite{Egge99}, \cite{Shy99} and \cite{Putter05}. Alternatively, \cite{2005_YangPrentice} presented a semiparametric two-sample model (hereafter denoted as YP model) for this type of problem. The feature ``two-sample'' refers to the scenario where, for example, there is a treatment and a control group that can be conveniently represented through a binary variable. The YP proposal is an interesting option, since it includes the PH and PO representations as particular cases. In their model, the baseline hazard function is left unspecified, in fact a counting process is assumed leading to a survival step function. A pair of short-term and long-term hazard ratio parameters is included to accommodate crossing survival curves. In addition, a pseudo maximum likelihood approach is considered for the estimation procedure. Consistency and asymptotic normality of the resulting estimators are demonstrated in the paper. 

\cite{2011_Yang} extended the estimation procedure in \cite{2005_YangPrentice} to pointwise and simultaneous inference on the hazard ratio function itself. They further proved the consistency and asymptotic normality of the estimates at a fixed time point. \cite{2012_Yang} proposed two omnibus tests to evaluate the adequacy of the YP model. The first test is based on the martingale residuals and the second one examines the contrast between the non-parametric and model-based estimators of the survival function. \cite{2013_Diao} extended the two-sample YP model to a general regression setting with possibly time-dependent covariates; the study developed an efficient likelihood-based estimation procedure. The authors also demonstrated the consistency, asymptotic normality and efficiency of the resulting estimators. The YP model has also been extended by \cite{2007_Tong} to accommodate current status survival data. Another extension is found in \cite{2017_Zhang} to fit case II interval-censored data.

The use of semiparametric methods for univariate survival data started with \cite{1972_Cox} on the proportional hazards model. \cite{Bres72} and \cite{Bres74} are two initial publications proposing  the use of the piecewise exponential (PE) distribution to replace the baseline hazard in a survival analysis. The grid configuration for a PE model is a central topic in \cite{Kalb73}; they explore different interval sizes for regular grids. According to the authors, the grid should be chosen independently of the data. Many applications, related to clinical trials and involving the PE distibution, can be found in the literature; some few examples are: leukemia \citep{Bres74}, gastric cancer \citep{Gam91}, kidney infection \citep{Sahu97,Ibra01}, breast cancer \citep{Sinha99}, melanoma \citep{Dem14} and hospital mortality \citep{Cla02}. Although parametric in a strict sense, the PE model has a strong nonparametric appeal. The main reason is the fact that assumptions about the shape of the baseline hazard are not required in this approach. 

The main contribution of the present paper is to propose a novel fully likelihood-based approach to handle right-censored survival data with crossing survival curves. This is done by assuming the PE distribution to deal with the baseline hazard in the YP model. We emphasize that using the semiparametric PE approach to extend the original YP model has never been considered in the literature. Some important advantages of the methodology proposed here are: ($i$) it has the tractability of parametric models; ($ii$) it provides a continuous survival function being convenient for the detection of the intersection point of two survival curves; ($iii$) it has the flexibility of a semiparametric model allowing different shapes for the hazard function, in contrast with the limited counting process assumed in the YP model and their extensions; ($iv$) the routine for maximum likelihood estimation and inference is straightforward and easy-to-implement. Another point to be highlighted is the fact that the original reference for the YP model is focused on the two-sample case leaving the general regression setting for future work. We explore the PE model with categorical and continuous covariates in this paper.   

This work is organized as follows. The proposed model is described in Section \ref{sec_model}. A comprehensive Monte Carlo simulation study is conducted in Section \ref{sec_sim} to evaluate the performance of the models. Section \ref{sec_applic} shows an empirical illustration where the new model is applied to study the survival times of patients enrolled in a gastric cancer clinical trial. Finally, Section \ref{sec_conclusions} presents the main conclusions, final remarks and discuss future research.

\section{Model formulation} \label{sec_model}

Let $T$ be a nonnegative random variable representing the time until the occurrence of an event of interest. In order to accommodate survival data with crossing survival curves, \cite{2005_YangPrentice} proposed the following model
\begin{eqnarray}
\label{St_yp}
 S(t|\bz) = \left[ 1 + \frac{ \lambda }{ \theta }R_{0}(t)\right]^{-\theta},
\end{eqnarray}
where $\bz=(z_{1}, \cdots, z_{p})$ is a set of explanatory variables, $\lambda=\exp\{\bz \bpsi\}$ and $\theta=\exp\{\bz \bphi\}$, $\bpsi=(\psi_{1}, \cdots, \psi_{p})$ and $\bphi=(\phi_{1}, \cdots, \phi_{p})$ are vectors of regression coefficients without intercepts, and $R_{0}(t)=F_{0}(t)/S_{0}(t)$ corresponds to the baseline odds.

The hazard function, associated with \eqref{St_yp}, can be expressed as
\begin{eqnarray}
\label{ht_yp}
 h(t|\bz) =  \frac{\lambda\theta}{\lambda F_{0}(t)+\theta S_{0}(t)}h_{0}(t),
\end{eqnarray}
where $F_{0}(t) = 1 - S_{0}(t)$ and $h_{0}(t) = -\frac{d}{dt}\log\left(S_{0}(t)\right)$.

The YP model has some interesting and attractive features. First, it is easy to see from \eqref{St_yp} and \eqref{ht_yp} that the PH and PO models arise as particular cases when $\bpsi=\bphi$ and $\bpsi=\mathbf{0}$, respectively. Another point is that a scenario with crossing survival curves can be obtained when $\psi_{j}\phi_{j} < 0$, for any pair of coefficients ($\psi_{j}$,$\phi_{j}$) and $j = 1, \cdots, p$. Finally, it follows from \eqref{ht_yp} that 
\begin{eqnarray}
  \lim_{t \rightarrow 0} \frac{h(t|\bz)}{h(t|\mathbf{0})} = \lambda ~~~ \mbox{and} ~~~ \lim_{t \rightarrow \infty}\frac{h(t|\bz)}{h(t|\mathbf{0})} = \theta
\end{eqnarray}

The quantities $\lambda$ and $\theta$ can be interpreted as the short-term and long-term hazard ratios, respectively. In addition, the elements $\bpsi$ and $\bphi$ can be regarded as the short-term and long-term regression coefficients, respectively. 

We now describe the main aspects related to the piecewise exponential distribution. Consider a time grid $\rho=\{a_{1},..., a_{m-1}\}$ inducing the following set of intervals:
\begin{equation}
    I_{k}=\left\{ 
    \begin{array}{l}             
      (a_{k-1}, a_{k}], ~~k=1, ..., m-1 \\
      (a_{m-1}, \infty), ~~k=m                  
    \end{array}\right. ,
\end{equation}
with $a_{0}=0$. We shall assume that the baseline hazard function appearing in \eqref{ht_yp} is constant in each interval induced by $\rho$, that is
\begin{equation}
    h_{0}(t|\bxi,\rho)=\xi_{k},
\end{equation}
for $t \in I_{k}$ and $k = 1, \cdots, m$.

The choice of the time grid in $\rho$ has a significant impact in terms of goodness-of-fit for the target model. A time grid with a large number of intervals might provide unstable estimates for the failure rates. On the other hand, time grids with few intervals might lead to poor approximations to the true survival function. In practice, the time grid selection must seek a balance in terms of how well the hazard and survival functions can be estimated. Several approaches have been proposed in the literature to address this issue. We shall assume here that the time grid $\rho$ is a known quantity composed by a subset of the observed failure times. For a detailed discussion regarding the choice of $\rho$, we recommend reading \cite{2011_Demarqui} and references therein. 

Following \cite{2011_Demarqui}, the baseline survival function $S_{0}(t|\bxi, \rho)$ can be conveniently expressed as:
\begin{equation}
\label{St}
    S(t|\bxi, \rho) = \exp\left\{ -\sum_{k=1}^{m}\xi_{k}(t_{k}-a_{k-1})\right\},
\end{equation}
where 
\begin{equation}
\label{def_tj}
t_{k}=
  \left\{\begin{array}{ll}
  a_{k-1}, & \mbox{if $t < a_{k-1}$} \\
  t, & \mbox{if $t \in I_{k}$} \\
  a_{k}, & \mbox{if $t > a_{k}$,} \\
\end{array}\right.  \nonumber
\end{equation}
for $k=1, ..., m$.

Consider now a random sample of size $n$, all elements are independent, and denote by $T_{i}$ and $C_{i}$ the failure and censoring times, respectively. Let $\bz_{i}$ be a $1\times p$ vector of explanatory variables associated with the $i$-th element in the sample. Assume that the censoring mechanism is non-informative. In addition, the failure times are right-censored so that $Y_{i}=\min\{T_{i}, C_{i}\}$ is the observable failure time. The term $\delta_{i}=I\{T_{i} \leq C_{i}\}$, for $i = 1, \cdots, n$, is the failure indicator function. The set of observed data is then denoted by $D=\{(y_{i}, \delta_{i}, \bz_{i}); \ i = 1, \cdots, n\}$. Finally, let $\bTheta=(\bpsi, \bphi, \bxi)$ represent the set of parameters to be estimated. Since the time grid $\rho$ is regarded as a known quantity in this paper, its notation will be suppressed here for simplicity. 

The likelihood function can be expressed as follows:
\begin{eqnarray} \label{lik}
 L(\bTheta;D) = \prod_{i=1}^{n}\left[ \frac{\lambda_{i}\theta_{i} } {\theta_{i} S_{0}(y_{i}|\bxi)+ \lambda_{i} F_{0}(y_{i}|\bxi)}h_{0}(y_{i}|\bxi) \right]^{\delta_{i}}\left[ 1+\frac{\lambda_{i}}{\theta_{i}}R_{0}(y_{i}|\bxi) \right]^{-\theta_{i}}, 
\end{eqnarray}
where $F_{0}(y_{i})=1-S_{0}(y_{i})$, $\lambda_{i}=\exp\{\bz_{i} \bpsi\}$ and $\theta_{i}=\exp\{\bz_{i} \bphi\}$.

In order to obtain the maximum likelihood estimates (MLEs) for the parameters, denoted by $\hat{\bTheta}$, we proceed to the direct maximization of the log-likelihood function $l(\bTheta) = \log L\left( \bTheta; D \right)$ by using the quasi-Newton BFGS method available in standard statistical softwares such as \texttt{R} \citep{softR} and \texttt{SAS} (\url{www.sas.com}). The BFGS has been widely used in the literature to solve optimization problems; see \citep{fle2000} for details. Finally, the variance-covariance matrix of $\hat{\bTheta}$ can be approximated by inverting the observed information matrix $\mathscr{I}(\hat{\bTheta})=-l''(\hat{\bTheta})$, which is readily provided, if requested, when applying the BFSG through the \texttt{R} general purpose optimization command \texttt{optim}.

In the next section we empirically investigate some asymptotic properties of the MLEs through a simulation study. This is a comprehensive study based on synthetic data replicated in a Monte Carlo (MC) scheme. The main idea is to explore different aspects of the proposed model and compare its results with those from the standard YP model.

\section{Simulation study} \label{sec_sim}

In this section, we present a Monte Carlo simulation study to evaluate the performance of the model introduced in the previous section. There are two main purposes in this analysis: ($i$) compare the proposed model with the two-sample semiparametric model in \cite{2005_YangPrentice} and ($ii$) evaluated the performance of the new model in the general regression setting. 

In order to generate the synthetic data sets, the Weibull baseline survival function $S_{0}(t|\alpha,\gamma)=\exp\left\{-\gamma t^{\alpha}\right\}$, with $\alpha = 1.50$ and $\gamma = 0.05$, is assumed to generate the failure times ($t_i$'s). The censoring times ($c_i$'s) are obtained from the $\mbox{U}(0,\tau)$, with $\tau$ chosen so that the censoring rate is approximately $30\%$ of the observed data. Recall that the final time reported for each sample unit is given by $y_{i} = \min\{t_{i}, c_{i}\}$. We begin the simulation study with the two-sample scenario. The MC schemes are configured with $1{,}000$ data sets and they explore three different sample sizes: $n=50$, $n=100$ and $n=200$. In each case, a single binary covariate is included assuming $z_{i} \sim \mbox{Bernoulli}(0.5)$, for $i = 1, \cdots, n$. 

All models were implemented and fitted using the \texttt{R} programming language \citep{softR}. In terms of optimization, we emphasize that the \texttt{BFGS} method is applied through the function \texttt{optim} available in \texttt{R}. The semiparametric YP model in \cite{2005_YangPrentice} can be fitted through the \texttt{R} package \texttt{YPmodel}; see more details in \cite{2010_Yang}, \cite{2011_Yang} and \cite{2012_Yang}.

The survival function, defined in the semiparametric YP model, is a step function with jumps on the observed failure times. In order to ensure a fair comparison between our PE model and the original YP model, the endpoints of the intervals forming the grid in the PE model are set to be the observed failure times. In other words, each interval contains exactly $1$ observation. Naturally, other configurations including more than $1$ time point per interval can be applied and this is expected to improve results.

\begin{table}[ht]
\caption{Summary for the MC simulation study with $1{,}000$ replications and a single binary covariate. Notation: fitted model (Mod), parameter name (Par), true value (True), average point estimate (Est.), average standard error (ASE), sample standard deviation of the estimates (SSDE), relative bias (RB), average 95\% confidence interval and coverage probabilities (CP).}
\centering \small
\begin{tabular}{cccccccccc} \hline 
 & \multicolumn{9}{c}{n=50} \\ \hline
 & & & & & & & \multicolumn{2}{c}{95\% CI} \\ \cline{8-9}
Mod   & Par & True & Est. & ASE & SSDE & RB(\%) & Lower & Upper & CP \\  \hline
PE  & $\psi$ & 1.0 & 0.907 & 0.858 & 0.870 & -9.269 & -0.773 & 2.588 & 0.940 \\ 
  & $\phi$ & -1.0 & -0.651 & 3.363 & 1.788 & 34.872 & -7.242 & 5.940 & 0.981 \\ \cline{2-10}
YP & $\psi$ & 1.0 & 1.150 & 1.419 & 1.522 & 14.976 & -1.631 & 3.931 & 0.863 \\ 
  & $\phi$ & -1.0 & -0.729 & 1.068 & 1.147 & 27.118 & -2.822 & 1.364 & 0.924 \\ \hline
 & \multicolumn{9}{c}{n=100} \\ \hline
 & & & & & & & \multicolumn{2}{c}{95\% CI} \\ \cline{8-9}
Mod   & Par & True & Est. & ASE & SSDE & RB(\%) & Lower & Upper & CP \\  \hline
PE  & $\psi$ & 1.0 & 0.955 & 0.595 & 0.606 & -4.511 & -0.212 & 2.122 & 0.947 \\ 
  & $\phi$ & -1.0 & -0.935 & 0.387 & 0.384 & 6.513 & -1.694 & -0.176 & 0.969 \\ \cline{2-10}
YP & $\psi$ & 1.0 & 1.138 & 2.042 & 1.098 & 13.787 & -2.864 & 5.140 & 0.946 \\ 
   & $\phi$ & -1.0 & -0.930 & 1.701 & 0.585 & 6.992 & -4.265 & 2.405 & 0.993 \\  \hline      
 & \multicolumn{9}{c}{n=200} \\ \hline
 & & & & & & & \multicolumn{2}{c}{95\% CI} \\ \cline{8-9}
Mod   & Par & True & Est. & ASE & SSDE & RB(\%) & Lower & Upper & CP \\  \hline
PE  & $\psi$ & 1.0 & 0.991 & 0.417 & 0.418 & -0.944 & 0.173 & 1.808 & 0.949 \\ 
  & $\phi$ & -1.0 & -0.966 & 0.258 & 0.265 & 3.434 & -1.471 & -0.461 & 0.956 \\ \cline{2-10}
YP  & $\psi$  & 1.0 & 1.039 & 3.007 & 0.590 & 3.939 & -4.854 & 6.933 & 0.994 \\ 
  & $\phi$ & -1.0 & -0.939 & 2.412 & 0.530 & 6.135 & -5.666 & 3.789 & 0.995 \\    \hline  
\end{tabular}
\label{tab_pe_versus_yp}
\end{table}

The relative bias reported in Table \ref{tab_pe_versus_yp} is calculated according to the following formulation:
$$
 RB(\kappa) = 100 \ (\hat{\kappa} - \kappa_{\tiny \mbox{true}}) \ / \ |\kappa_{\tiny \mbox{true}}|.
$$
In this expression consider that: $\kappa$ is a generic parameter, $\hat{\kappa}$ is the maximum likelihood estimate and $\kappa_{\tiny \mbox{true}}$ is the true value. The relative bias is basically the ratio between the estimation error and the magnitude of the true value. Negative and positive results indicate underestimation and overestimation, respectively. The fraction is multiplied by $100$ to adjust scale leading to a quantity indicating a percentage representing how big is the error with respect to the magnitude of the true value. This quantity is commonly used in the survival analysis literature. 

Table \ref{tab_pe_versus_yp} shows the results of the Monte Carlo simulation study. As can be seen, neither the proposed model nor the YP model performed well when the sample size is small ($n=50$). In this case, both models show relative biases above $\approx 10\%$ and coverage probabilities far from the nominal level for all parameters. Now looking at the moderate sample sizes ($n=100$ and $n=200$) the results in Table \ref{tab_pe_versus_yp} change in favour of the proposed PE model. It is evident that the PE model has a superior performance with respect to the standard semiparametric YP model. Although an improvement in terms of relative bias reduction can be observed for both models under moderate sample sizes, the proposed model provides smaller relative biases and indicates coverage probabilities closer to the nominal level of $95\%$. 

Another important aspect exhibited in Table \ref{tab_pe_versus_yp} is the similar results for the ASE and SSDE related to the proposed PE model. The similarity between these quantities is expected in a MC study, since the MC error for a parameter tend to reflect, on average, the estimator standard error in each replicated data set. Note that this type of result is not true for the standard semiparametric YP model, which seems to overestimate the standard errors of the parameter estimators. In addition, this bad behavior can explain the wider average $95\%$ confidence interval limits and the coverage probabilities above the nominal level observed for this model.

We now turn our attention to the general regression setting. Moderate to large data sets were considered in this analysis to investigate the performance of the proposed PE model assuming a regression structure with four covariates. Synthetic data sets were simulated taking into account three different sample sizes: $n=100$, $n=200$ and $n=500$. We again use the MC scheme with $1{,}000$ replications. The following short-term and long-term linear predictors are explored:
\begin{equation}
 \begin{array}{l} 
 \log(\lambda_{i}) = +2.0z_{1i} - 0.5z_{2i} + 1.5z_{3i} - 1.5z_{4i}  \\ 
 \log(\theta_{i})  = -1.0z_{1i} + 1.0z_{2i} - 1.5z_{3i} + 1.5z_{4i}
 \end{array} \label{lpred}
\end{equation}
where $z_{1i} \sim \mbox{Bernoulli}(0.5)$, $z_{2i} \sim \mbox{N}(0, 1)$, $z_{3i} \sim \mbox{Bernoulli}(0.5)$ and $z_{4i} \sim \mbox{N}(0, 1)$, for $i = 1, \cdots, n$.

The main reference \cite{2005_YangPrentice} is entirely focused on the two-sample scenario and does not explore the general regression setting. In fact the paper indicates that the standard YP model can be extended to incorporate covariates, but this was left for future work in that opportunity. The corresponding \texttt{R} package \texttt{YPmodel}  does not allow the analysis using the general configuration. As a consequence of this point, in the next analysis we do not confront the results from the PE model and the standard YP case. Recall that, for comparison reasons, the time grid for the PE model was initially chosen (analysis of Table \ref{tab_pe_versus_yp}), with $1$ observation per interval. The results presented in Table \ref{tab_reg} are obtained by assuming a different grid structure. In this case, the number of intervals is given by $m = \sqrt{n}$. This choice is convenient to reduce the computational burden to fit the model. The endpoints of the intervals are chosen according to the ideas described in \cite{2011_Demarqui}.

\begin{table}[h]
\caption{Summary for the MC simulation study with $1{,}000$ replications and $4$ covariates. Notation: parameter name (Par), true value (True), average point estimate (Est.), average standard error (ASE), sample standard deviation of the estimates (SSDE), relative bias (RB), average 95\% confidence interval and coverage probabilities (CP).} 
\centering \footnotesize
\begin{tabular}{ccccccccc}   \hline
 \multicolumn{9}{c}{$n=100$ and $m=10$} \\ \hline
   & & & & & & \multicolumn{2}{c}{95\% CI} \\ \cline{7-8}
  Par & True & Est. & ASE & SSDE & RB(\%) & Lower & Upper & CP \\  \hline
  $\psi_{1}$ & 2.0 & 1.903 & 0.614 & 0.656 & -4.855 & 0.700 & 3.106 & 0.931 \\ 
  $\psi_{2}$ & -0.5 & -0.473 & 0.296 & 0.315 & 5.314 & -1.053 & 0.106 & 0.940 \\ 
  $\psi_{3}$ & 1.5 & 1.415 & 0.615 & 0.655 & -5.657 & 0.209 & 2.621 & 0.919 \\ 
  $\psi_{4}$ & -1.5 & -1.477 & 0.402 & 0.416 & 1.562 & -2.264 & -0.689 & 0.942 \\ \hline
  $\phi_{1}$ & -1.0 & -0.913 & 0.347 & 0.397 & 8.664 & -1.593 & -0.234 & 0.927 \\ 
  $\phi_{2}$ & 1.0 & 1.047 & 0.238 & 0.258 & 4.738 & 0.581 & 1.514 & 0.951 \\ 
  $\phi_{3}$ & -1.5 & -1.459 & 0.350 & 0.372 & 2.718 & -2.145 & -0.773 & 0.952 \\ 
  $\phi_{4}$ & 1.5 & 1.547 & 0.266 & 0.295 & 3.156 & 1.026 & 2.068 & 0.964 \\ \hline
 \multicolumn{9}{c}{$n=200$ and $m=15$} \\ \hline
   & & & & & & \multicolumn{2}{c}{95\% CI} \\ \cline{7-8}
  Par & True & Est. & ASE & SSDE & RB(\%) & Lower & Upper & CP \\  \hline
  $\psi_{1}$ & 2.0 & 1.946 & 0.422 & 0.412 & -2.704 & 1.119 & 2.773 & 0.945 \\ 
  $\psi_{2}$ & -0.5 & -0.479 & 0.200 & 0.209 & 4.181 & -0.871 & -0.087 & 0.937 \\ 
  $\psi_{3}$ & 1.5 & 1.430 & 0.423 & 0.438 & -4.692 & 0.600 & 2.259 & 0.933 \\ 
  $\psi_{4}$ & -1.5 & -1.475 & 0.276 & 0.284 & 1.689 & -2.015 & -0.935 & 0.937 \\ \hline
  $\phi_{1}$ & -1.0 & -0.953 & 0.233 & 0.237 & 4.657 & -1.411 & -0.496 & 0.955 \\ 
  $\phi_{2}$ & 1.0 & 1.034 & 0.160 & 0.162 & 3.367 & 0.720 & 1.348 & 0.945 \\ 
  $\phi_{3}$ & -1.5 & -1.477 & 0.237 & 0.251 & 1.535 & -1.941 & -1.012 & 0.939 \\ 
  $\phi_{4}$ & 1.5 & 1.528 & 0.179 & 0.185 & 1.848 & 1.178 & 1.878 & 0.937 \\ \hline
  \multicolumn{9}{c}{$n=500$ and $m=23$} \\ \hline
   & & & & & & \multicolumn{2}{c}{95\% CI} \\ \cline{7-8}
  Par & True & Est. & ASE & SSDE & RB(\%) & Lower & Upper & CP \\  \hline
  $\psi_{1}$ & 2.0 & 1.976 & 0.264 & 0.270 & -1.223 & 1.458 & 2.493 & 0.936 \\ 
  $\psi_{2}$ & -0.5 & -0.485 & 0.123 & 0.124 & 2.905 & -0.727 & -0.244 & 0.948 \\ 
  $\psi_{3}$ & 1.5 & 1.463 & 0.264 & 0.261 & -2.441 & 0.946 & 1.981 & 0.949 \\ 
  $\psi_{4}$ & -1.5 & -1.472 & 0.171 & 0.174 & 1.866 & -1.807 & -1.137 & 0.936 \\ \hline
  $\phi_{1}$ & -1.0 & -0.978 & 0.144 & 0.143 & 2.177 & -1.261 & -0.695 & 0.958 \\ 
  $\phi_{2}$ & 1.0 & 1.010 & 0.098 & 0.097 & 1.035 & 0.818 & 1.202 & 0.953 \\ 
  $\phi_{3}$ & -1.5 & -1.482 & 0.146 & 0.142 & 1.218 & -1.769 & -1.195 & 0.955 \\ 
  $\phi_{4}$ & 1.5 & 1.518 & 0.110 & 0.115 & 1.227 & 1.303 & 1.734 & 0.931 \\ \hline
 \end{tabular}
\label{tab_reg}
\end{table}

As it can be seen from Table \ref{tab_reg}, relative biases are reasonably low, especially for $n=200$ and $n=500$. In addition, the coverage probabilities are, in general, close to the nominal level of $95\%$. Another important aspect observed here is the fact that both bias and ASE tend to decrease as the sample size increases. The results displayed in Table \ref{tab_reg} also indicate that the standard errors of the parameters are being well estimated, since the ASE and SSDE have similar values for all parameters; this is true regardless of the sample size under investigation. Overall, the proposed model seems to perform well in the general regression setting for moderate to large data sets.

\section{Real data application} \label{sec_applic}

This section is dedicated to the analysis of a real data set freely available through the \texttt{R} package \texttt{YPmodel} under the label of \texttt{gastric}; see also \cite{1982_GTSG} as a formal reference for more details. This gastric cancer data set has become a common application in the literature related to survival analysis and, more specifically, it can be easily found in studies dealing with crossing survival curves; some few references are: \cite{2012_Yang}, \cite{2013_Diao}, \cite{2011_Lee} and \cite{2018_Yang}. The experiment in this clinical trial involves $90$ individuals diagnosed with locally unresectable (advanced) gastric cancer. The participants were randomly assigned to the following groups: ($i$) the control group composed by $45$ patients receiving chemotherapy and ($ii$) the treatment group including $45$ patients receiving a combination of chemotherapy and radiation therapy. These individuals were followed within this study for about $5$ years. Three variables are reported in the data set for each patient: the time response representing either a failure (time to death) or a right censoring, a binary failure indicator identifying those patients experiencing the event of interest and, finally, a group binary indicator with $1$ meaning the treatment category. Note that this application contains a single binary covariate; therefore, it can be explored and compared via the PE and YP standard models.   

\begin{figure}[!h] 
\centering
 $$
 \begin{array}{cc}\\
  \includegraphics[scale=0.27]{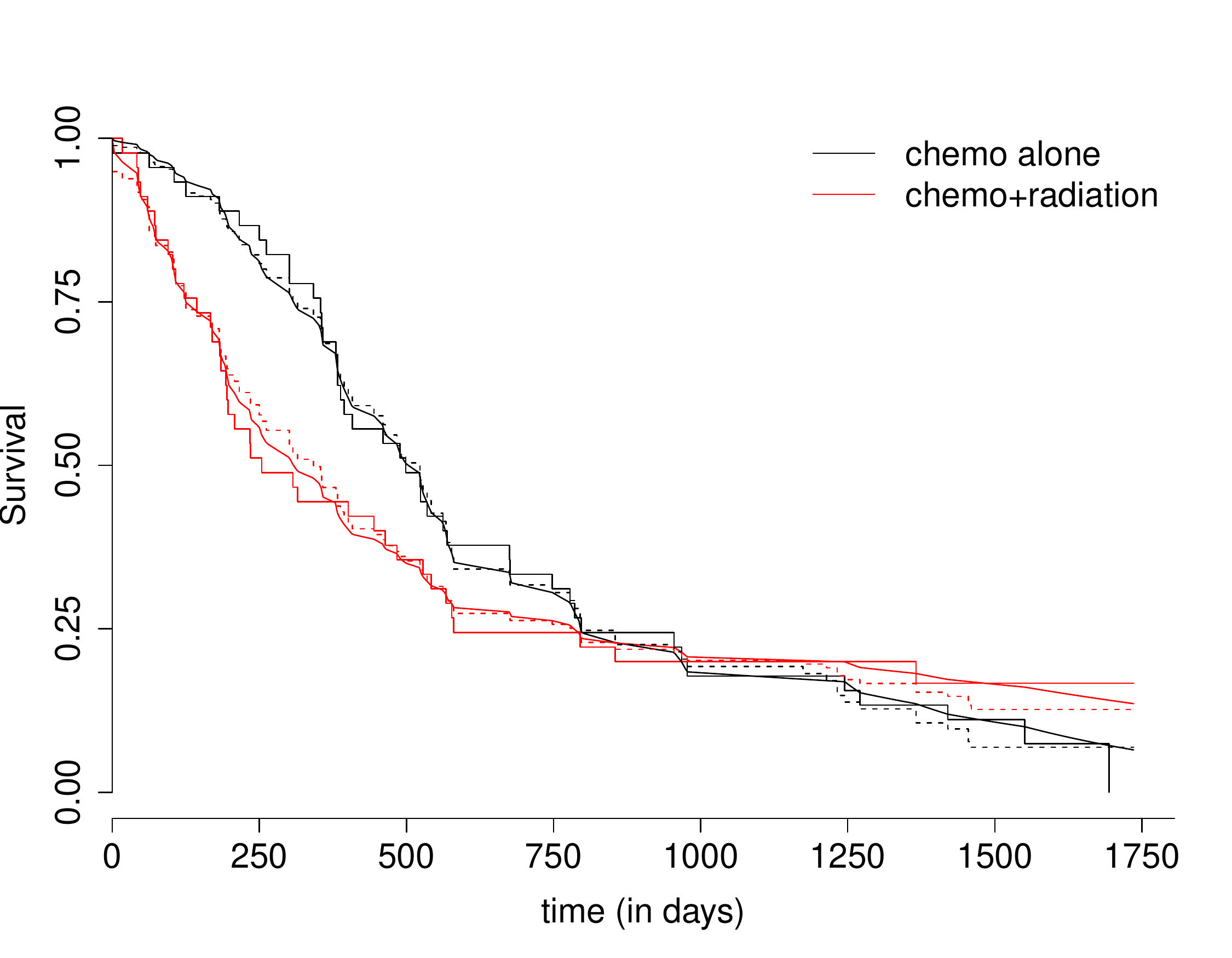} &
  \includegraphics[scale=0.27]{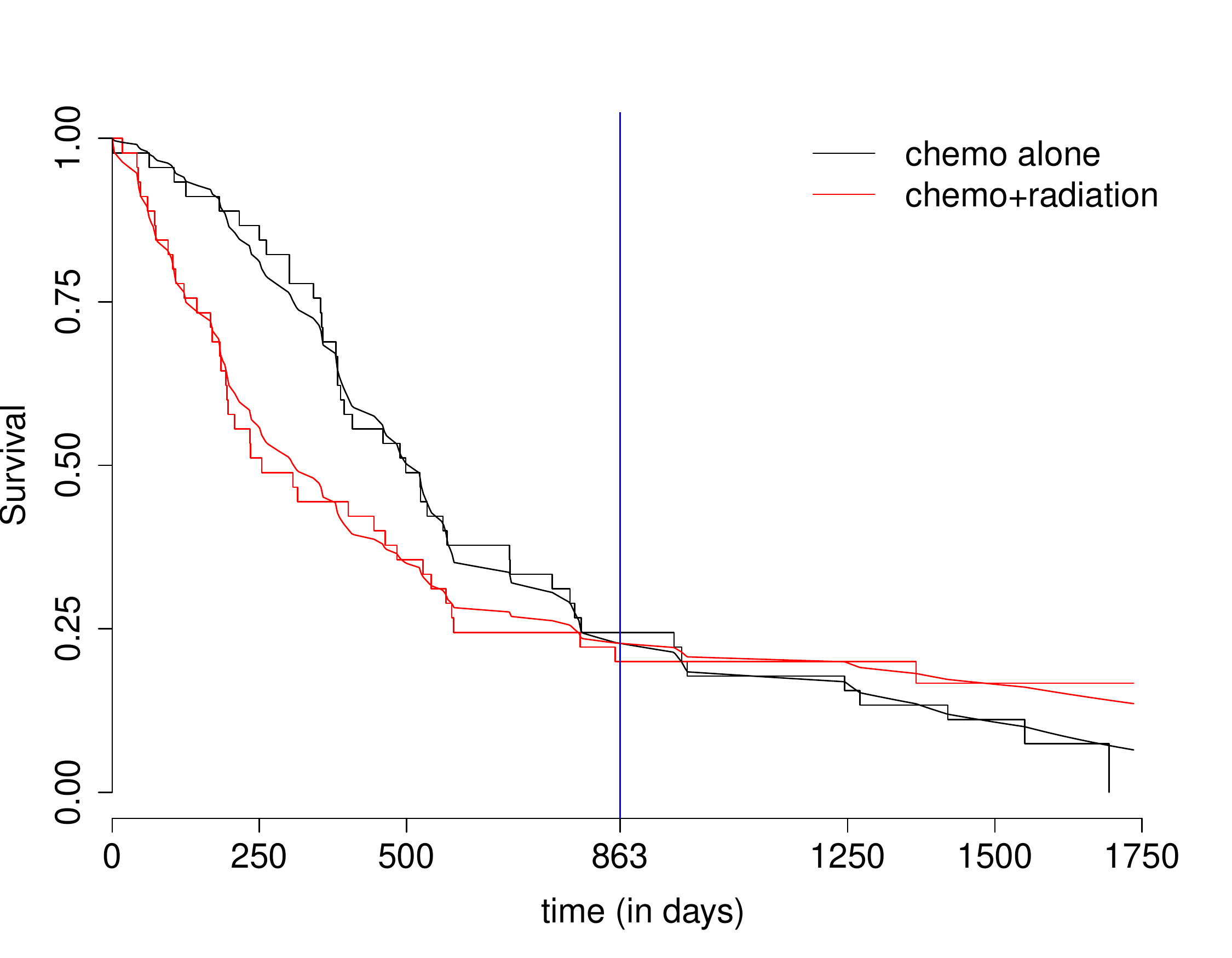} \\ 
 \end{array}
 $$ 
\caption{Analysis of the gastric cancer data set. Left panel: Kaplan-Meier estimates for the survival curves along with the PE estimates (solid lines) and the YP estimates (dashed lines). Right panel: estimated survival curves provided by the proposed model along with the estimated time point at which the survival curves probably cross each other.}
\label{fig_St}
\end{figure}

Table \ref{tab_gastric} summarizes the results obtained for both models. As it can be observed, the short-term ($\psi$) and long-term ($\phi$) regression coefficients, within each model, are estimated with opposite signs and they have distinct magnitudes. This can be observed by either looking at the point estimate (column Est.) or the $95\%$ confidence intervals. This behavior is a clear indication of survival curves having an intersection at some intermediate time point between $0$ and the maximum. In other words, the top and bottom positioning of the curves are inverted for the intervals below and above the crossing time point; see Figure \ref{fig_St} for a visual idea. This inversion suggests the existence of an alteration in the effectiveness of the treatment at some point during the follow up period of the study. In general, the results tend to be similar when comparing the corresponding estimates from both models. Note that the standard error related to $\psi$ is larger then the one for $\phi$. In addition, all p-values from the z-test are small, indicating significant estimates.     

\begin{table}[ht]
\centering
\caption{Summary of the models fitted to the gastric cancer data.}
\begin{tabular}{cccccccc}   \hline
& & & & \multicolumn{2}{c}{95\% CI} \\ \cline{5-6}
 Model & Par & Est. & SE & lower & upper & z & p-value \\    \hline
PE & $\psi$ & 1.837 & 0.648 & 0.567 & 3.108 & 2.834 & 0.005 \\ 
   & $\phi$ & -1.017 & 0.300 & -1.606 & -0.429 & -3.387 & 0.001 \\ \hline
YP & $\psi$ & 1.600 & 0.538 & 0.547 & 2.656 & 2.977 & 0.003 \\ 
   & $\phi$ & -0.906 & 0.248 & -1.393 & -0.421 & -3.650 & 0.000 \\ 
   \hline
\end{tabular} \label{tab_gastric}
\end{table}

One interesting and attractive feature of the proposed PE model is the fact that it provides a continuous survival function. This aspect allows us to apply standard procedures to find the roots of nonlinear equations to determine accurately the time point at which the survival curves intersect each other. One possibility to handle this problem in \texttt{R} is to use the command \texttt{uniroot} \citep{Brent73} for unidimensional searches. In line with this idea, right after fitting our PE model to the gastric cancer data, we apply the \texttt{uniroot} function to conclude that the crossing time occurs, for this application, at the time point given by the day $863$ within the full period of the study. From the practical point of view, this means that before the day $863$, the patients in the control group (only chemotherapy) have  better survival rates than those in the treatment group. On the other hand, the benefits of combining chemotherapy with radiotherapy tend to emerge in a later stage of the study (after day $863$).

The left panel of Figure \ref{fig_St} shows: the Kaplan-Meier curves associated with the two treatments and the survival curves estimated via the semiparametric PE and YP models. The right side panel of Figure \ref{fig_St} displays the estimated survival curves, provided by the proposed PE model, along with the estimated time at which the survival curves are expected to cross. As can be seen in the left panel, both models seem to accommodate and represent well the data, since their estimated survival curves tend to agree with the Kaplan-Meyer survival estimates. 

\section{Conclusions} \label{sec_conclusions}

This paper presents a fully likelihood-based approach to deal with crossing survival curves as an extension to the standard YP model proposed in 2005 for a two-sample case. The main difference with respect to other extensions of the YP model is the fact that we take advantage of the piecewise exponential semiparametric modeling to allow a flexible representation of the baseline hazard function. This also configures the main contribution of the paper, since no other study combining these two aspects (YP model structure and PE distribution) can be found in the literature of survival analysis. Using the PE distribution brings some advantages when comparing to other semiparametric options for the YP model. The PE model preserves the flexibility of the semiparametric models and the tractability of the parametric ones. In addition, it is relatively easy-to-implement using standard maximization routines. Estimation of parameters, hazard function, survival function and hazard ratios is straightforward. Another important aspect to be emphasized is the fact that the survival function has a continuous representation via the PE model; this is not true in the original YP model and other approaches presented in the literature, where a step function is obtained as the survival representation. As a result of this feature, the time in which the survival curves (treatment and control groups) intersect each other can be easily and accurately determined.

A comprehensive MC simulation study was developed to examine the performance of the proposed PE model in comparison with the YP model. The results indicate that the PE model provides better results with smaller relative biases being observed for most parameters. Using synthetic data sets, the behavior of the PE model was also investigated for a general regression setting involving several covariates. The standard YP model can be extended to this context, but the original paper in 2005 does not explore this type of result. Our findings suggest that the PE model also has a good performance when dealing with several covariates.

The final analysis of the paper is focused on a real application related to a well known data set related to a clinical trial for patients detected with advanced gastric cancer. In summary, the results of the PE and YP model are similar and they clearly indicate significant regression coefficients with opposite signs, which is expected for the scenario where the survival curves have an intersection.  

In terms o future work, we plan to develop an \texttt{R} package to fit the proposed PE model. The approach presented here can also be extended to accommodate survival data with cure fraction and interval-censored observations. This is beyond the scope of the current paper and will be addressed in upcoming papers.

\vspace{10pt}

\noindent {\large \bf Acknowledgements} \vspace{5pt}

The second author gratefully acknowledge the support from Funda\c{c}\~{a}o de Amparo a Pesquisa do Estado de Minas Gerais (FAPEMIG). 

\bibliographystyle{biom}
\bibliography{references}

\begin{thebibliography}{}

\bibitem[\protect\citeauthoryear{Brent}{Brent}{1973}]{Brent73}
Brent, R.~P. (1973).
\newblock {\em Algorithms for minimization without derivatives}.
\newblock Prentice-Hall, Englewood Cliffs, 1 edition.

\bibitem[\protect\citeauthoryear{Breslow}{Breslow}{1972}]{Bres72}
Breslow, N. (1972).
\newblock Discussion on regression models and life-tables (by {D. R. Cox}).
\newblock {\em Journal of the Royal Statistical Society, Series B} {\bf 34,}
  216--217.

\bibitem[\protect\citeauthoryear{Breslow}{Breslow}{1974}]{Bres74}
Breslow, N. (1974).
\newblock Covariance analysis of censored survival data.
\newblock {\em Biometrics} {\bf 30,} 89--99.

\bibitem[\protect\citeauthoryear{Clark and Ryan}{Clark and Ryan}{2002}]{Cla02}
Clark, D.~E. and Ryan, L.~M. (2002).
\newblock Concurrent prediction of hospital mortality and length of stay from
  risk factors on admission.
\newblock {\em Health Services Research} {\bf 37,} 631--645.

\bibitem[\protect\citeauthoryear{Cox}{Cox}{1972}]{1972_Cox}
Cox, D.~R. (1972).
\newblock Regression models and life-tables.
\newblock {\em Journal of the Royal Statistical Society, Series B} {\bf 34,}
  187--220.
\newblock with discussion.

\bibitem[\protect\citeauthoryear{Demarqui, Dey, Loschi, and Colosimo}{Demarqui
  et~al.}{2011}]{2011_Demarqui}
Demarqui, F.~N., Dey, D.~K., Loschi, R.~H., and Colosimo, E.~A. (2011).
\newblock {\em Modeling Survival Data Using the Piecewise Exponential Model
  with Random Time Grid}, pages 109--122.

\bibitem[\protect\citeauthoryear{Demarqui, Dey, Loschi, and Colosimo}{Demarqui
  et~al.}{2014}]{Dem14}
Demarqui, F.~N., Dey, D.~K., Loschi, R.~H., and Colosimo, E.~A. (2014).
\newblock Fully semiparametric {B}ayesian approach for modeling survival data
  with cure fraction.
\newblock {\em Biometrical Journal} {\bf 56,} 198--218.

\bibitem[\protect\citeauthoryear{Diao, Zeng, and Yang}{Diao
  et~al.}{2013}]{2013_Diao}
Diao, G., Zeng, D., and Yang, S. (2013).
\newblock Efficient semiparametric estimation of short-term and long-term
  hazard ratios with right-censored data.
\newblock {\em Biometrics} {\bf 69,} 840--849.

\bibitem[\protect\citeauthoryear{Egge and Zahl}{Egge and Zahl}{1999}]{Egge99}
Egge, K. and Zahl, P.~H. (1999).
\newblock Survival of glaucoma patients.
\newblock {\em Acta Ophthalmologica Scandinavica} {\bf 77,} 397--401.

\bibitem[\protect\citeauthoryear{Fletcher}{Fletcher}{2000}]{fle2000}
Fletcher, R. (2000).
\newblock {\em Practical methods of optimization}.
\newblock John Wiley and Sons, New York, 2 edition.

\bibitem[\protect\citeauthoryear{Gamerman}{Gamerman}{1991}]{Gam91}
Gamerman, D. (1991).
\newblock Dynamic {B}ayesian models for survival data.
\newblock {\em Journal of the Royal Statistical Society, Series C} {\bf 40,}
  63--79.

\bibitem[\protect\citeauthoryear{{Gastrointestinal Tumor Study
  Group}}{{Gastrointestinal Tumor Study Group}}{1982}]{1982_GTSG}
{Gastrointestinal Tumor Study Group} (1982).
\newblock A comparison of combination chemotherapy and combined modality
  therapy for locally advanced gastric carcinoma.
\newblock {\em Cancer} .

\bibitem[\protect\citeauthoryear{Ibrahim, Chen, and Sinha}{Ibrahim
  et~al.}{2001}]{Ibra01}
Ibrahim, J.~G., Chen, M.~H., and Sinha, D. (2001).
\newblock {\em {B}ayesian survival analysis}.
\newblock Springer series in statistics. Springer-Verlag, New York.

\bibitem[\protect\citeauthoryear{Kalbfleisch and Prentice}{Kalbfleisch and
  Prentice}{1973}]{Kalb73}
Kalbfleisch, J.~D. and Prentice, R.~L. (1973).
\newblock Marginal likelihoods based on {C}ox's regression and life model.
\newblock {\em Biometrika} {\bf 60,} 267--278.

\bibitem[\protect\citeauthoryear{Lee}{Lee}{2011}]{2011_Lee}
Lee, S.~H. (2011).
\newblock Maximum of the weighted {Kaplan-Meier} tests for the two-sample
  censored data.
\newblock {\em Journal of Statistical Computation and Simulation} {\bf 81,}
  1017--1026.

\bibitem[\protect\citeauthoryear{Putter, Sasako, Hartgrink, van-de Velde, and
  van Houwelingen}{Putter et~al.}{2005}]{Putter05}
Putter, H., Sasako, M., Hartgrink, H.~H., van-de Velde, C. J.~H., and van
  Houwelingen, J.~C. (2005).
\newblock Long-term survival with non-proportional hazards: results from the
  {D}utch gastric cancer trial.
\newblock {\em Statistics in Medicine} {\bf 24,} 2807--2821.

\bibitem[\protect\citeauthoryear{{R Core Team}}{{R Core Team}}{2018}]{softR}
{R Core Team} (2018).
\newblock {\em R: A Language and Environment for Statistical Computing}.
\newblock R Foundation for Statistical Computing, Vienna, Austria.

\bibitem[\protect\citeauthoryear{Sahu, Dey, Aslanidou, and Sinha}{Sahu
  et~al.}{1997}]{Sahu97}
Sahu, S.~K., Dey, D.~K., Aslanidou, H., and Sinha, D. (1997).
\newblock A {W}eibull regression model with gamma frailties for multivariate
  survival data.
\newblock {\em Lifetime Data Analysis} {\bf 3,} 123--137.

\bibitem[\protect\citeauthoryear{Shyur, Elsayed, and Luxhoj}{Shyur
  et~al.}{1999}]{Shy99}
Shyur, H.~J., Elsayed, E.~A., and Luxhoj, J.~T. (1999).
\newblock A general model for accelerated life testing with time-dependent
  covariates.
\newblock {\em Naval Research Logistics} {\bf 49,} 303--321.

\bibitem[\protect\citeauthoryear{Sinha, Chen, and Ghosh}{Sinha
  et~al.}{1999}]{Sinha99}
Sinha, D., Chen, M.~H., and Ghosh, S.~K. (1999).
\newblock {B}ayesian analysis and model selection for interval-censored
  survival data.
\newblock {\em Biometrics} {\bf 55,} 585--590.

\bibitem[\protect\citeauthoryear{Tong, Zhu, and Sun}{Tong
  et~al.}{2007}]{2007_Tong}
Tong, X., Zhu, C., and Sun, J. (2007).
\newblock Semiparametric regression analysis of two-sample current status data,
  with applications to tumorigenicity experiments.
\newblock {\em Canadian Journal of Statistics} {\bf 35,} 575--584.

\bibitem[\protect\citeauthoryear{Yang, , and Zhao}{Yang
  et~al.}{2012}]{2012_Yang}
Yang, S., , and Zhao, Y. (2012).
\newblock Checking the short-term and long-term hazard ratio model for survival
  data.
\newblock {\em Scandinavian Journal of Statistics} {\bf 39,} 554--567.

\bibitem[\protect\citeauthoryear{Yang}{Yang}{2018}]{2018_Yang}
Yang, S. (2018).
\newblock Improving testing and description of treatment effect in clinical
  trials with survival outcomes.
\newblock {\em Statistics in Medicine} {\bf 38,} 530--544.

\bibitem[\protect\citeauthoryear{Yang and Prentice}{Yang and
  Prentice}{2005}]{2005_YangPrentice}
Yang, S. and Prentice, R.~L. (2005).
\newblock Semiparametric analysis of short-term and long-term hazard ratios
  with two-sample survival data.
\newblock {\em Biometrika} {\bf 92,} 1--17.

\bibitem[\protect\citeauthoryear{Yang and Prentice}{Yang and
  Prentice}{2010}]{2010_Yang}
Yang, S. and Prentice, R.~L. (2010).
\newblock Improved logrank-type tests for survival data using adaptive weights.
\newblock {\em Biometrics} {\bf 66,} 30--38.

\bibitem[\protect\citeauthoryear{Yang and Prentice}{Yang and
  Prentice}{2011}]{2011_Yang}
Yang, S. and Prentice, R.~L. (2011).
\newblock Estimation of the 2-sample hazard ratio function using a
  semiparametric model.
\newblock {\em Biostatistics} {\bf 12,} 354--368.

\bibitem[\protect\citeauthoryear{Zhang, Wang, and Sun}{Zhang
  et~al.}{2017}]{2017_Zhang}
Zhang, H., Wang, P., and Sun, J. (2017).
\newblock Regression analysis of interval-censored failure time data with
  possibly crossing hazards.
\newblock {\em Statistics in Medicine} {\bf 37,} 768--775.

\end{thebibliography}

\end{document}